\documentclass{aa}
\usepackage{graphicx}
\usepackage{txfonts}
\usepackage{natbib}
\bibpunct{(}{)}{;}{a}{}{,} 

\begin{document}
 \title{The evolution of the AGN content in groups  up to z$\sim$ 1}
  \author{L. Pentericci\inst{1}, M. Castellano\inst{1}, N. Menci\inst{1},
    S. Salimbeni\inst{2}, T. Dahlen\inst{3},  A. Galametz\inst{1}, P. Santini\inst{1},
    A. Grazian\inst{1} and A. Fontana\inst{1} }
\offprints{Laura Pentericci pentericci@oa-roma.inaf.it}
 \institute{INAF - Osservatorio Astronomico di Roma, Via Frascati 33,
I--00040, Monte Porzio Catone, Italy
\and Astronomy Department, University of Massachusetts, Amherst, MA,
  01003, USA
\and Space Telescope Science Institute
3700 San Martin Drive Baltimore, MD 21218 USA}
\date{}
\abstract{We explore the AGN content in  groups in the two
  GOODS fields (North and South), exploiting the ultra-deep 2 and 4 Msec Chandra
  data and the deep multiwavelength observations from optical to mid IR 
  available for both fields.}
{ Determining the AGN content in structures of different
  mass/velocity dispersion and comparing them to higher 
mass/lower redshift analogs
   is important to understand how the AGN formation process  is related to
   environmental properties.}{We use our well-tested cluster finding algorithm to
identify structures in the two GOODS fields, exploiting the available
spectroscopic redshifts as well as accurate photometric redshifts. We identify 9
 structures in GOODS-south (already presented in a previous paper) and 8 new
structures in the GOODS-north field. We only consider structures where at least
2/3 of the members brighter than $M_R=-20 $ have a spectroscopic
redshift.
We then check if any of the group members 
coincides with X-ray sources that belong to  the 4 and 2 Msec source catalogs
respectively, and
with a simple classification based on total rest-frame hard luminosity and 
hardness ratio we determine 
if the X-ray emission originates from AGN activity or it is more probably related to
the galaxies' star-formation activity.}
{We find that the fraction of AGN with $ Log L_H > 42 erg s^{-1}$ in galaxies with 
$M_R < -20$ varies from less than $5\%$ to $22\%$ with an average value of $6.3 \pm 1.3\%$, i.e. much
higher than the value found for lower redshift groups of similar mass, which is just
1\%. It is also more than double the fraction found for massive
clusters at a similar high redshift ($z\sim 1$). 
We then  explore the spatial distribution of
 AGN in the structures and find that they preferentially  populate the outer
regions rather than the  center. The colors of AGN host galaxies 
in structures tend to be confined to the green valley, thus avoiding the blue cloud and
partially also the red-sequence, contrary to what happens in the field.
We finally compare our results to the predictions of two sets of semi analytic
models to investigate the evolution of AGN and evaluate potential triggering
and fueling mechanisms. The outcome of this comparison attests the
importance  of galaxy encounters, not 
necessarily leading to mergers, as an efficient AGN triggering mechanism.
}{}
\keywords{}
\authorrunning{L. Pentericci et al.}
\titlerunning{The evolution of the AGN content in groups and small clusters up to z$\sim$ 1}
\maketitle
\section{Introduction}
The frequency  and properties of active galactic nuclei (AGN) in the field, 
groups  and clusters can provide  information about how these objects are
triggered and fueled. In fact,  one of the possible mechanisms that trigger AGN activity 
is the interaction and merging of galaxies \citep{Barnes1996}, 
enabling the creation of a central super-massive black hole and the 
matter to fuel it. In this context  the AGN fraction would be heavily 
influenced  by the environment  providing the opportunities 
for interaction and the supply  of fuel, and should therefore 
strongly depend on the local density. 
In addition to a comparison between the AGN fraction in different 
environments, measurements of the evolution of the AGN population in clusters 
 can constrain the formation time of their super-massive 
black holes and the extent of their co-evolution with the cluster galaxy
population   \citep{Martini2009}. 
Indeed  the external conditions are likely to heavily depend on the cluster evolutionary
stage, and can be very different in structures that have recently merged  
compared to   massive virialized clusters \citep{VanBreukelen2009}. 

X-ray observations are 
essential in the study of active galaxies since a considerable fraction of
X-ray  selected AGN do not show in their spectra the emission lines 
characteristic of optically selected AGN \citep{Martini2002}. 
This suggests the existence of a large population of obscured, or at least
optically unremarkable AGN. 
This result  is attributed to the higher sensitivity  of X-ray observations to
lower-luminosity AGN relative to visible-wavelength emission-line diagnostics.
In particular, deep X-ray observations can  probe also the relatively faint AGN population, 
associated to more ``normal'' galaxies and not just to the extremely massive ones.
The Chandra Deep Field North and South are currently the areas
with the deepest available X-ray observations, having
 a total of 2 and 4 Ms of data respectively. They are therefore 
the ideal locations to study AGN with moderate luminosity up to relative 
high redshifts.  Although the AGN population in the CDFS  has been extensively
studied \cite[e.g.][]{Mainieri2005,Trevese2007},  
our project is focused on  the association between AGN
(relatively faint ones) with  groups  and small clusters that have been detected 
in the two fields at intermediate redshifts. 
Both fields were the subjects of very extensive observational campaigns 
at practically all wavelengths, from optical to near and mid-IR 
(including deep Spitzer data). Last but not least, about 2000 spectra 
were obtained on each area from several groups
\citep{Vanzella2006,Vanzella2008,Popesso2009,Balestra2010}.

In this paper we will assess the fraction of AGN in  groups
from  z$\sim 0.5$ to z$\sim 1.1$. The paper is organized as follows: 
in Section 2 we present the detection of the
structures using a 3D algorithm based on photometric redshifts. 
In Section 3 we present the identification of group  members with the X-ray
sources; in Section 4   we determine the fraction of AGN in each of our
structures and discuss the dependence of this fraction on both redshift and 
velocity dispersion, using  complementary data from the literature on lower
redshift/more massive systems. We also determine the colors and spatial
distribution of AGN. Finally in Section 5 we compare our results to the
prediction of different semi-analytic models and discuss their implications.
\\
Throughout the paper all magnitudes are in the AB system, and we adopt
$H_0=70$~km/s/Mpc, $\Omega_M=0.3$ and $\Omega_{\Lambda}=0.7$.

\section{Structures and groups in GOODS North and South fields}
\subsection{Detection}
It has been shown \citep{Eisenhardt2008,Salimbeni2009,VanBreukelen2009} 
that high quality photometric redshifts can be effectively used to 
 find and study clusters at redshift above
 1, where X-ray detection
techniques  become progressively less efficient, due to surface brightness
dimming and SZ surveys are only just beginning to give 
preliminary detections \citep{Vanderlinde2010}. Other methods 
rely on assumptions that are not necessarily fulfilled at these 
early epochs, such as the 
presence of a well defined red sequence \citep{Gladders2000,Andreon2009}.

\begin{table*}
\begin{tabular}{ccccccc}

ID   &  z  &  RA & Dec      &M200(b=1/2)    &  R200(b=1/2)& $N_{spec}$ \\ 
     &     & J2000 & J2000  & $M_\odot$   & Mpc   &  \\
\hline 
GN 1    & 0.638    & 189.0283  &  62.1711 & 2.7/1.4 E+14 &   1.47/1.66 & 16   \\ 
GN 2    & 0.484    & 189.1686  &  62.2152  & 2.3/0.6 E+14 &  1.53/0.94 & 21   \\ 
GN 3    & 1.014    & 189.1589  &  62.1860  & 3.5/1.1 E+14 &  1.29/0.89 & 16  \\ 
GN 4    & 0.863    & 189.1271  &  62.1476  & 8.1/3.2 E+13 &  0.87/0.67 & 22  \\
GN 5    & 0.851    & 189.1783  &  62.2777  &4.4/2.0 E+14  &  1.52/1.20 & 37  \\ 
GN 6    & 1.014    & 189.2089  &  62.3304  & 9.7/4.5 E+13&  0.88/0.67 & 9   \\ 
GN 7    & 0.973    & 189.3422  &  62.1918  & 2.7/1.3 E+14 &  1.24/0.97 & 9    \\ 
GN 8    & 0.457    & 189.4867  &  62.2595  & 1.0/0.5 E+14 &  1.25/0.97 & 13  \\ 
\hline
\end{tabular}
\caption[]{Characteristics of groups and clusters in GOODS-North }
\end{table*}
In this context,  we have developed the ``(2+1)D algorithm''
providing an adaptive estimate of the 3D density field,
using positions and photometric redshifts \citep{Trevese2007} that
 can be used in an efficient way to detect candidate galaxy clusters
and groups. On the basis of accurate simulations we have shown that
our algorithm can individuate groups and clusters with a very low
spurious detection rate and a high completeness up to redshift $\sim
2$ \citep[for a detailed  description of these simulations see Sect. 3 in][]{Salimbeni2009}.
This algorithm has been  extensively applied to the  GOODS-North and South
fields (Giavalisco et al. 2004), where extremely accurate 
photometric redshifts can be determined  thanks to the deep multiwavelength 
photometry available in many bands \citep{Grazian2006}. 
In particular the $z_{850}$-selected catalogue of the GOODS-South field
includes
photometric redshifts for $\sim$10000 galaxies with an r.m.s. $\Delta
z/(1+z)\sim 0.03$ up to redshift 2 \citep{Santini2009}. The 
GOODS-North field includes photometric redshifts for $\sim$10000 galaxies 
with an r.m.s. $\Delta z/(1+z)\sim 0.045$ up to redshift 2 (Dahlen et al. in prep).
\\
Despite the fact that  the areas studied are not very large (each
field is approximately $10'\times 15 '$ for a total of about 300 arcmin$^2$), and therefore we
do not expect to find  rare massive clusters, we  identify  several structures, that we characterize 
as  groups and small  clusters.
Indeed, one of the most distant clusters known to date, CL0332-2742 at z=1.61 
was found by our group using this algorithm \citep{Castellano2007} 
and was then  spectroscopically confirmed with independent follow up 
observations by the GMASS collaboration \citep{Kurk2009}.

\subsection{Groups and  clusters characteristics}
In the GOODS-south  field we find several  structures  up to z$\sim 2$ 
that have been extensively described in \citet{Salimbeni2009}.
Of these, two are classified as small clusters  and the rest as groups based on
the masses derived from the galaxy over-density and/or from the velocity
dispersion. 
We will consider only the  structures up to  redshift$\sim$1, for consistency with the  GOODS-North field
where the  larger photometric redshift uncertainty does not allow us
 to reach a similar accuracy at z$> 1$.
\\
In the GOODS-North field we  find 8 structures 
 up to redshift $\sim 1$. 
In Table 1  we report the groups and cluster characteristics derived from the 
algorithm, namely the peak position of the over-density, the mean redshift. We
report the mass determined from the over-density value and the radius,
assuming a bias parameter 1 and 2. In particular, the mass $M_{200}$ is 
defined as the mass inside the radius corresponding to a density contrast
$\delta_m =delta_{gal}/b \sim 200$ (Carlberg et al. 1997), where b is the bias 
factor (see Salimbeni et al. 2009 for more details). In the Table we also
report 
the number of spectroscopically confirmed galaxies.
Briefly, of the new  structures in GOODS-north, CIG1236+6215 (GN 5) at z=0.85 
was originally identified by \cite{Dawson2001} with 8 spectroscopic members
and was then
 reported by \cite{Bauer2002} as a possibly under-luminous X-ray
cluster, using the then available 1 Msec Chandra observation.
We now assign 37  spectroscopic members to this cluster.
While nobody specifically reported on the other structures in the 
GOODS-north field,  \citet{Barger2008} and Elbaz et al. (2007) both  noticed the presence of large scale
structures at z=0.85 and z=1 from the spectroscopic redshift distribution.

In this work, we  restrict our analysis to 
the  structures that have a large fraction of
member galaxies with accurate spectroscopic redshifts,  
the main reason being  the need to determine the total number of group/cluster members to derive 
the AGN fraction as accurately as possible.
Specifically we select groups/clusters that have an accurate spectroscopic
redshift for at least 65\%  of the  members 
brighter than  $M_R < -20$  (the limit that will be used to determine the AGN
fraction). 
In total, 5 structures from GOODS-South and 6 from GOODS-North comply with
this requirement.
This does not mean that the other structures are unreal, but only that the 
fraction of AGN determined would be more  uncertain due to the unknown 
 number of real bright cluster members.
\\
We use all spectroscopic galaxies  (including in some cases objects 
with a magnitude below the considered limit) to  
determine the structure spectroscopic center and the velocity 
dispersion: we apply a clipping in velocity of $\pm 2000 km s^{-1}$ from
 the center.
If the redshift is farther than 2000 $km s^{-1}$  from the spectroscopic 
center, the galaxy 
 is considered as an interloper. The number of interlopers is typically very low (1-4 per structure).
For groups containing less than 15 spectroscopic members,  the velocity dispersion  
is derived using the  Gapper sigma statistics \citep{beers}, while for the others we use the
 normal statistics.
The dispersions obtained  are  in the range  370 to 640 km $s^{-1}$ 
and the masses are of the order of 0.5 to few times $10^{14} M\odot$ \citep{Salimbeni2009}:
these values confirm  that we are observing structures that range from  groups to small-sized  clusters.  
\\
It is not trivial to evaluate the  status of  our groups and clusters 
as fully formed  and virialized structures, given that  
for some we only have few spectroscopic redshift.
For the two most massive structures, where the
number of redshifts is sufficiently high, we assessed two of the most
important cluster characteristics, i.e. the presence of the 
red-sequence and the virialization status.
 In Figure 1 (lower panel) we present the color magnitude relation for  the most
  massive structure in GOODS-North (GN 5) for both spectroscopic and
  photometric cluster members. The presence of the  red-sequence
    is clear.  To check if the  cluster has reached a relaxed status (virial
equilibrium) we analyse the velocity distribution of the spectroscopic
members. Indeed this status, which is acquired through the process of “violent
relaxation” (Lynden-Bell 1967), is characterised by a Gaussian galaxy velocity
distribution (e.g. Nakamura 2000) and, as shown by
N-body simulations, by a low mass fraction included in substructures
(e.g. Shaw et al. 2006). 
In the upper left panel of Figure 1 we show the  binned velocity distribution
of the spectroscopic members, compared to Gaussians with dispersion obtained
through the  biweight estimate (red) and considering the jackknife uncertainties (blue and
green lines). 
We then performed five one-dimensional statistical tests to investigate whether the
velocity distribution of the galaxy members is consistent with being Gaussian:
the Kolmogorov-Smirnov test (as implemented in the ROSTAT package of Beers et
al. 1990), two classical normality tests (skewness and kurtosis) and the two
more robust asymmetry index (A.I.) and tail index (T.I.) described in Bird \&
Beers (1993). We find consistency with a Gaussian in all cases.
We then performed the two-dimensional Δ-test of Dressler \& Shectman (1988) to
look for substructures and found no evidence.
\\
The observed color magnitude diagram and the results of the 
tests for Gaussianity and substructures
for one of the most massive structure in GOODS-South (GS 4)
have   been presented and discussed in Castellano et al. (2011). In that 
paper there are also additional  details on the tests performed. 
For the other structures it is not possible to carry out such tests since the
number of spectroscopic members is too low to give meaningful results.

\begin{figure}
\includegraphics[width=9.5cm,clip=]{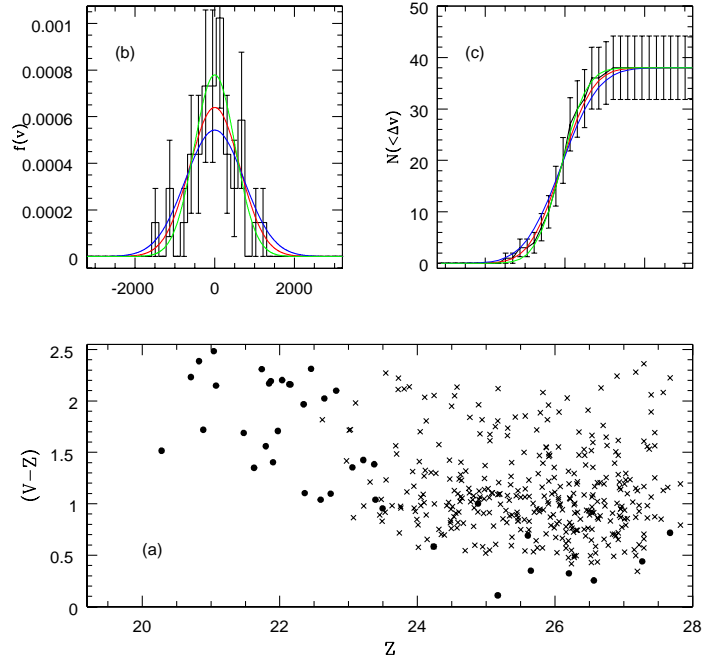}
\caption{ Upper left
panel: binned velocity distribution of the spectroscopic members, compared to
Gaussians with dispersion obtained through the biweight estimate (red) and
considering the jackknife uncertainties (blue and green). All distributions
are normalized to 1.0. Upper right panel: cumulative velocity distributions,
colour code as in the left panel. 
Lower panel: observed colour magnitude diagram of GN 5 of all
  spectroscopic (black circles) and photometric 
(crosses) members of the cluster within 1 Mpc from the center. }
\label{fig:fig1}
\end{figure}

\begin{figure}
\includegraphics[width=9cm,clip=]{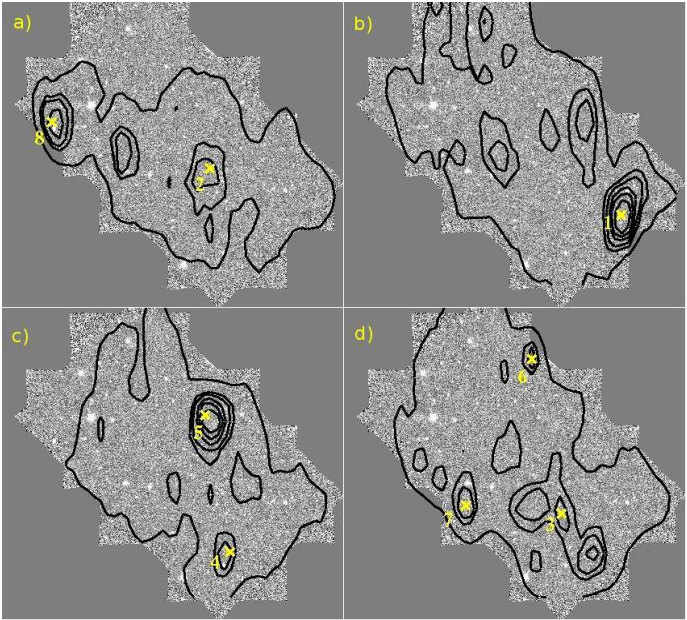}
\caption{Density isosurfaces for structures at $z\sim 0.45-0.48$ a), at $z\sim 0.64$ b), $z\sim 0.85$ c) and at $z\sim 0.97-1.01$ d) (average, average $+2\sigma $, average $+3\sigma $ to average $+10 \sigma $) superimposed on the ACS z850 band image of the GOODS-North field. Yellow crosses indicate the density peak of each structure, the number is the ID of the structure in Table 1.}
\label{fig:fig2}
\end{figure}
Figure 2 shows the density isosurfaces for the structures in GOODS-North 
superimposed on the ACS z850 band images of the field. Figure 3 
 shows the positions of the  overdensities over the photometric
  redshift distribution of the entire GOODS-North sample. The overdensities
  are also traced by the  distribution of the spectroscopically confirmed AGN
  in our catalogue, as shown in the lower panel of this Figure (note that AGN are not included in the 
sample used for the density estimation ). 
The analogous figures for GOODS-South were presented in Salimbeni et al. (2009).

\subsection{X-ray emission from the clusters and groups}
An inspection of the Chandra images at each cluster/group 
position shows that only two of the structures have  significant extended 
X-ray emission due to the hot IGM. 
These  are cluster GS 5  and cluster
GN 8.  The emission from  Cluster GS 5 can be modeled with a  Raymond Smith
model  
with a best fit temperature 2.6 keV  and metallicity 0.2 $Z_\odot$: the
resulting  X-ray luminosity   is $9.56 \times 10^{42} erg s^{-1}$ in 
the 0.1-2.4 keV rest-frame band (Castellano et al. in preparation).
For GN 8 we can not estimate a temperature from the data: in this case the 
X-ray luminosities in the 0.1-2.4 keV rest-frame band is  4.1$\times 10^{42}
erg s^{-1}$, assuming a Raymond Smith model  with temperature 1 keV  and metallicity 0.2 $Z_\odot$.
All other structures, including the most massive ones,
are undetected: as argued by \citet{Salimbeni2009}, this lack of X-ray emission   
possibly indicates that optically selected  structures are X-ray under-luminous, at least when
  compared to X-ray selected ones. 
This is for example the case of GS 4 (or CIG  0332-2747)  at z=0.734, which was extensively discussed in 
Castellano et al. (2011), where we showed also  a tentative $\sim 3 \sigma $
detection of the X-ray emission, corresponding to a luminosity of $2 \times
10^{42} erg s^{-1}$.
\\
To further check the reality  of our groups  we performed  a stack of the
X-ray emission for all the new structures found in GOODS-North. 
We first measured the count rates in the soft band, within a square of
 side of $\sim $ $30 \hbox{$^{\prime\prime}$ }$   centred on the position of
 the peak  of each structure, as  given by our algorithm.
This aperture was used to be consistent with Salimbeni et al. (2009)
and corresponds approximately to a 1Mpc radius 
(of course depending slightly  on redshift), which is similar to the $R_{200}$ reported in Table 1.
We then masked all X-ray sources present within this area.
We finally subtracted the  soft-band background which was calculated  
from  the total exposure map, by  taking the
total integration time at the position 
of each group and multiplying it by the average background count rate of 0.056
counts Ms$^{-1}$ pixel${-1}$ (Alexander et al. 2003).  
Alternatively for each group we calculated the
average background count-rate in an annulus around the source  where no other 
sources were present. The two values in general agreed to within 1\%.
\\
For the combination of the 7 groups/clusters in GOODS-North that
are individually undetected (all but GN 8) we get an average
of 310$\pm 60$ counts  ($\sim 5.2 \sigma$); if we only include the 5
groups that are used in this work, the result is 220$\pm 50$ ($\sim 4.4
\sigma$).
We convert the measured count rate  to rest frame total $L_x$
 in the 0.1–-2.4 keV band, assuming  a metallicity $Z = 0.3   Z_\odot$  and a 
temperature kT = 1  keV. 
This temperature is typical for low redshift groups  with similar
 velocity dispersion (e.g. Osmond \& Ponman 2004).
We obtain a  luminosity of the order 1-2 $10^{42} erg s^{-1}$.
Compared to the typical luminosities of X-ray selected groups
with similar velocity dispersion in the local
Universe, the value we have found is on the low side, but still within
 the range of  the X-ray luminosities of these structures  
(Osman \& Ponman 2004). \\
We conclude that given the low mass of most of our structures and their 
high redshift, the lack of significant X-ray emission  is still 
consistent in most cases with the $L_x - \sigma$
relation, especially if one considers the larger scatter 
that is found for optically
selected structures (e.g. Rykoff et al. 2008).
We  caution that, although the results from the X-ray
stacking are encouraging, it is impossible with the present data to test the  
  virialization status of the individual groups.

\begin{figure}
\includegraphics[width=9cm,clip=]{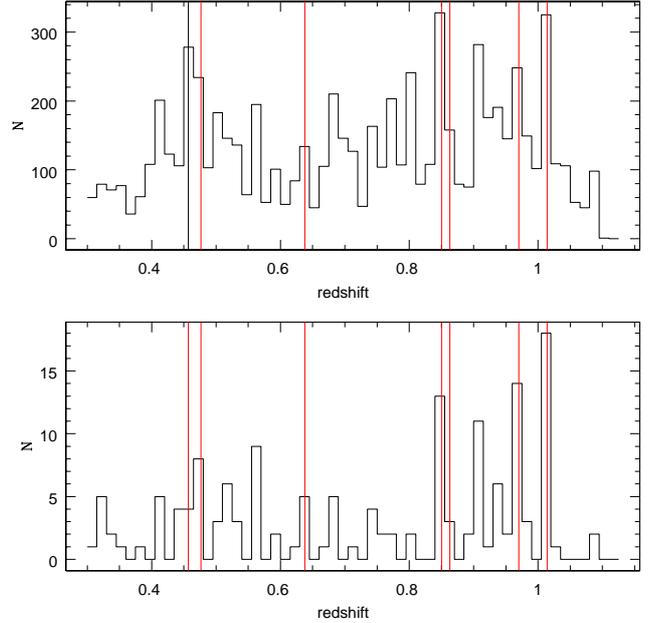}
\caption{Upper panel: photometric redshift distribution of our sample
  (continuous line). The vertical red lines mark the redshifts of the detected
  structures. 
Lower panel: redshift distribution of spectroscopically selected AGN in the GOODS-North }
\label{fig:AGN}
\end{figure}

\begin{table*}
\begin{tabular}{cccccccccccc}
ID    & RA  & Dec & redshift & $M_R$  &  Flux  & Flux & Flux& ${L_H}$  & HR &
Class & $log(f_x/f_o)$ \\ 
      & 2000&2000 &          &            & tot    & 0.5-2 keV&2-8keV&
      $10^{42} erg s^{-1}$ & \\
\hline
\hline
GOODS-South \\ 
\hline 
8355   &  53.1564  & -27.8108 & 0.665   & -22.28   & -6.47e-17&1.71e-17&-9.85e-17 & $<$0.19 & $<$0.17  & SB  & -2.4   \\
9633   &  53.1554  & -27.7915 & 0.667   & -22.68   & -6.73e-17 &2.10e-17&-9.59e-17 & $<$0.18 &$<$0.06 & SB  & -1.6 \\
7031  & 53.0939  & -27.8305 &0.734    & -19.5   & 4.00e-16 & 5.06e-17&3.29e-16 & 0.88  & 0.13$\pm$0.20  & AGN2 & 0.0 \\
8977  & 53.0734  & -27.8033 &0.734   & -19.56  & 1.80e-16 &-1.82e-17 & 2.39e-16& 0.63  & $>$0.42   & AGN2 & -0.66 \\
9589  & 53.0618  & -27.7940 &0.735    & -20.28  & 1.01e-16 & 2.87e-17&-1.54e-16 & 0.16  & $<$0.14  & AGN2 & -0.85 \\
9792  & 53.0751  &-27.7885  &0.734    & -23.55  & 1.58e-16 & 8.33e-17&-1.20e-16 & 0.24  &$<$-0.42  & SB & -1.2     \\
10995 & 53.0644  &-27.7754  &0.735    & -20.71  &-8.47e-17 & 2.73e-17&-1.31e-16 & $<$0.35  & $<$0.09  & SB  & -1.2 \\
11184 & 53.1207	 &-27.7732  &0.737    & -20.24 & 6.66e-17&-2.33e-17&-1.40e-16& 0.10  & 0.2$\pm 0.4$  & AGN2 & -0.22 \\
11248 & 53.1197  &-27.7723  &0.738    & -21.83  & 6.83e-16 & 2.66e-16&
4.20e-16& 1.12   &-0.40$\pm 0.06$  & AGN1 & -0.22 \\
11719 & 53.0769  &-27.7655  &0.738    & -22.23  & 1.75e-16
&-2.12e-17&2.14e-16& $>$ 0.57   & $>$ 0.32  & AGN2 & -0.82 \\
13053 & 53.0602  &-27.7491  &0.738    & -20.00   & 4.50e-16 & 8.47e-17&
3.70e-16& 1.00   &-0.02$\pm 0.2$   & AGN2 & 0.73 \\
13552 & 53.0941  &-27.7405  &0.738    & -22.92  & 4.86e-16 & 3.55e-17&
4.72e-16& 1.26   & 0.44$\pm 0.3$   & AGN2 & -1.4 \\
1837  & 53.0803  &-27.9017  & 0.964  & -23.71 & 4.2e-16   & 2.1e-16 & -2.8e-16&       &    & cluster \\
11085  & 53.0658 & -27.7749 &1.021    & -21.44  & 1.38e-16 &
4.72e-17&-1.61e-16& 0.47  &$<$-0.08 & AGN2 & -1.51 \\
11285  & 53.0509 & -27.7724 &1.033    & -22.38  & 2.60e-15 & 4.32e-16&
2.22e-15& 13.5 & 0.04$\pm$0.07 & AGN2 &0.23  \\
3920   & 53.0715 & -27.8724 &1.097    & -21.73  & 1.42e-14 & 4.98e-16&
1.41e-14& 99.7 & 0.69$\pm 0.06$ & AGN2 & 1.03 \\
\hline 
GOODS-North \\
\hline
1795    &  188.9940 &  62.1842 & 0.638  & -22.40 & 6.6e-16  & 1.20e-16&  5.10e-16& 0.97 & -0.10$\pm 0.25$ & SB  & -1.1 \\
2600    &  189.0136 &  62.1865 & 0.638  & -22.74 & 4.1e-16  & 1.40e-16& -3.20e-16& 0.45 & $<$ -0.34 & SB & -1.54 \\
3349    &  189.0276 &  62.1643 & 0.637  & -21.84 & 3.22e-15 &-6.00e-17& 3.30e-15 & 6.28 &  $>$ 0.74 & AGN2 &  0.36 \\
10013   &  189.1215 &  62.1796 & 1.013  & -22.39 & 2.31e-15 & 1.70e-16& 2.23e-15& 12.9 &  0.34$\pm 0.15$ & AGN2 & 0.57 \\ 
11584   &  189.1403 &  62.1684 & 1.016  & -23.09 & 8.60e-16 & 2.50e-16& 6.00e-16& 3.50 & -0.32$\pm 0.09$ & AGN1 & -0.57\\
10066   &  189.1220 &  62.2706 & 0.848  & -22.59 & 2.54e-15 & 1.20e-16& 2.59e-15& 9.79 &  0.51$\pm 0.18$  & AGN2 & -0.18 \\
12018   &  189.1453 &  62.2746 & 0.848  & -21.79 & 6.3e-16 & 4.00e-17& 6.10e-16& 2.31 &  0.41$\pm 0.38$  & AGN2 & -0.31 \\
13907   &  189.1657 &  62.2634 & 0.848  & -22.89 & 1.9e-16  & 9.00e-17&-1.70e-16& 0.42 & $<$-0.43 & SB & -0.55 \\
14867   &  189.1758 &  62.2627 & 0.857  & -22.88 & 2.46e-15 & 6.9e-16 & 1.82e-15& 7.07 & -0.22$\pm 0.07$  & AGN & 0\\ 
16314   &  189.1926 &  62.2577 & 0.851  & -22.12 & 5.6e-16  & 1.9e-16 & 3.40e-16& 1.30 & -0.42$\pm 0.10$ & AGN  &0.12 \\
17850   &  189.2096 &  62.3347 & 1.011  & -21.79 & 1.81e-15& 6.7e-16 & 1.10e-15& 6.36 & -0.45$\pm 0.05$ & AGN1 & 0.79  \\
19243   &  189.2230 &  62.3386 & 1.023  & -21.23 & 1.73e-15& 3.0e-17 & 1.75e-15&10.40 &  0.7$\pm 0.40$ &  AGN2 & 0.45 \\
29746   &  189.3410 &  62.1767 & 0.978  & -21.17 & 2.11e-15& 5.0e-17 & 2.11e-15&11.21 &  0.67$\pm 0.27$  & AGN2 & -0.15 \\
36782   &  189.4461 &  62.2756 & 0.440  & -22.08 & 4.2e-16& 9.0e-17 & 5.80e-16& 0.38 &  $<$0.05 & SB & -1.2   \\
38454   &  189.4950 &  62.2494 & 0.457  & -20.18 & 6.4e-16  & 2.0e-16 &4.70e-16& 0.43& -0.33$\pm 0.20$  & SB  & -0.66 \\
\hline\hline
\end{tabular}
\caption{X-ray sources associated to galaxies brighter than $M_B$=-20; the Xray properties of the sources have been derived from the Chandra
  Deep Field South 4-Megasecond Catalog 
  (http://heasarc.gsfc.nasa.gov/W3Browse/chandra/chandfs4ms.html) 
 and the  Chandra Deep Field North 2-Megasecond Catalog
  (http://heasarc.gsfc.nasa.gov/W3Browse/all/chandfn2ms.html). The tipical
  on-axis 3$\sigma$ flux limits are $3.2\times 10^{-17}$, $9.1\times 10^{-18}$, and $5.5\times
  10^{-17} erg cm^{-2} s^{-1}$ for the full, soft, and hard bands, respectively
for the GOODS-South sources; 
$7.1 \times 10^{-17}$, $2.5
  \times  10^{-17}$ and $1.4\times 10^{-16} ergs cm^{-2} s^{-1}$ for the full,
  soft and  hard bands for the GOODS-North sources.
We refer to   the linked table for individual flux uncertainties.}
\label{popgal}
\end{table*}
\section{X-ray point sources identification and classification} 
Given the sensitivity of the 2 Msec observations
with a typical total flux limit of 7 $\times 10^{-17} erg s^{-1}  cm^2$,  
we are able to detect AGN with $L_H> 10^{42}$  at all redshifts up to  1.1
(the most distant structure in the present study) also in the shallower
GOODS-North field.
We  cross correlate the group/cluster member lists  
with the Chandra deep field north and
south source catalogs derived respectively by \citet{Alexander2003} and  \citet{Luo2008}.
The cross correlation was performed using  a radius of 2 arcsec, which can be
considered the nominal relative uncertainty of the astrometric solution.
In all cases there is no ambiguity in the identification of the Chandra
X-ray source with its optical counterpart.  

\begin{table*}
\center{
\begin{tabular}{ccccccccc}
 Cluster   &   z &  $\sigma$ & $N_{spec}$ &  $N(M_R<-20)$ &  $N_{AGN}>10^{42}$ & $\%$ &$N_{AGN}>10^{43}$& $\%$  \\
\hline
GS 1     & 0.666  & 390g   & 9  & 9      &  0   & $<$13.2   & 0 & $<$11.0  \\
\hline 
GS 4     & 0.735  & 600cl  & 65 & 59     &  3   & 5.1 & 0 & $<$3.1\\
\hline   
GS 5     & 0.966  & 420g   & 12 & 20     &  0   & $<$9.2  & 0 & $<$9.2  \\
\hline     
GS 6     & 1.038  & 630g   & 11 & 22     &  1   & 4.5 & 1 &4.5  \\
\hline   
GS 8     & 1.098  & 510g  &  9 & 14     &  1   & 7.1 & 1 &7.1\\
\hline
GN 1     & 0.638  & 590g   & 16 & 14     &  1   &7.1 & 0 &$<$13.1  \\
\hline 
GN 3     & 1.014  & 440cl  & 16 & 17     &  2   &11.8 & 1 &5.9  \\
\hline 
GN 5     & 0.851  & 600cl  & 37 & 45     &  4   & 8.9 & 0 &$<$4.1\\
\hline 
GN 6     &1.014   & 370g   & 9  &  9     &  2   &22.2 & 1 &11.1  \\
\hline 
GN 7     &0.973   & 370g   & 9  & 12     &  1   & 8.3 & 1 &8.3\\
\hline 
GN 8    &0.457   & 570g   & 13 &  16    &  0   & $<$11.5  & 0 &$<$11.5  \\
\hline
Total    &        &        &    & 237    & 15   & 6.3 & 5 &2.1\\
\end{tabular}
\caption[]{Cluster properties and AGN fraction; the velocity dispersion 
$\sigma$ is in km s$^{-1}$: ``g''
  indicates that is was determined using gaussian statistics, while ``cl'' indicates  that it was computed using the Gapper sigma statistics.}
\label{popgal}}
\end{table*}

In Table 2 we present the group/cluster members with an
absolute magnitude $M_R < -20$ which  coincide with an X-ray source. AGN in
galaxies fainter than this limit are not considered for consistency with
previous works (e.g. Arnold et al. 2009).
We find a total of 31 sources that are also members of our
groups/clusters. In the Table  we report  their ID  from the 
GOODS-MUSIC catalog for the southern field (Santini et al. 2009) 
and from  Dahlen et al. (in
preparation)  for the northern field; the positions; the 
 spectroscopic redshift; the total hard (2-8 keV) and soft band
 (0.5-2 keV)  fluxes from the  catalogs; 
the derived hardness ratio HR=(H-S)/(H+S), where H and S are the soft
 and hard band counts;   the total  inferred luminosity in the rest-frame hard
 band (2-10 KeV) obtained extrapolating the observed hard band flux and
 assuming a power law with photon index $\Gamma =1.8$. For those AGN which are
 undetected in the hard band, but are detected in the total band,  this last
 value was used. For those few
 that are detected only in the soft band (and  have upper limits both in the
 total and in the hard-band), we infer an upper limit for the total
 rest-frame hard band luminosity.
Note that all these X-ray sources have a spectroscopic redshift.
\\
The characterization of sources is not entirely trivial since we are probing deep enough  X-ray luminosities
that some of the  detected X-ray emission might
 be due to star-burst   rather than related to AGN activity (especially in the
 deeper GOODS-South field).
We will employ a very simple classification according to the total luminosity
and to the hardness ratio of the X-ray emission.
X-ray sources are classified as type 2 AGN if they have a hardness ratio
$HR\ge -0.2 $, regardless of their total luminosity.
They are classified as Type 1 AGN if they have  $HS\le -0.2 $ but a total X-ray
luminosity exceeding $10^{42} erg s^{-1}$. Sources with lower luminosity and soft 
emission, are classified as star-burst galaxies.
In some case only a limit in available for the HR, so the classification
  becomes ambiguous: in this cases we further considered  the
  X-ray-to-optical ratio ($log(f_x/f_o)$), which is defined as the ratio between the
  total X-ray flux and the B band flux (as in  Georgakakis et al. 2004). AGN broadly have $log(f_x/f_o)$
  in the range -1 to 1, so if  $log(f_x/f_o )<-1$, sources are considered star-burst.
In Table 2 we report this ratio for all sources.
 We finally  checked, whenever available in the literature, the optical
  spectra of the X-ray sources, or a classification  based
  on these spectra  (e.g. Szokoly et al. 2004, Trouille et al. 2008, Mignoli
  et al. 2004). 
Most  of the sources are classified as emission line galaxies or high
excitation emitters. None of the sources we could check were classified as
broad line AGN.

In conclusion, of the total sample of X-ray sources, 9  are classified as star-burst galaxies, 
15 are Type 2 AGN,  5  are Type 1 AGN, and one is associated to the
diffuse emission from  the hot cluster gas (although there could be a
component associated to the BC galaxy).
In some cases the classification maybe border line, however the AGN that we will
use in the rest of the analysis (those with $L> 10^{42} erg s^{-1} cm^{-2}$) all have a
solid classification.

\section{Results and discussion}
\subsection{AGN fractions in clusters and groups}
We determine  the fraction $f_A$ of  AGN in clusters and groups, by dividing
the number of AGN (regardless of type) by the  total 
number of  members down to an absolute magnitude limit $M_R =-20$.
All AGN have a spectroscopic redshift, but  cluster/group  members include  
also some galaxies with only a photometric redshift. While it is 
possible that some of the galaxies included are interlopers, we also expect
that galaxies belonging  to the structure could be placed out of the
structures due to a wrong photometric redshift. We will assume that these two
effects more or less compensate each other; in any case since we include only
structures with at least 65\% of spectroscopic members, we estimate that 
this uncertainty is minimal.
\\
Another way to estimate the global structure population is from the velocity
dispersion, using  the correlation  between this quantity and 
the total number of galaxies within $R_{200}$ 
found by Koester et al. (2007) which is   
 $ln \sigma= 5.52 +0.31 ln N_{R200}$. 
We derive the $N_{200}$ using this relation and it is  
in general agreement  with the total number of
cluster/group members derived from our photometric plus spectroscopic redshifts. 
\\
In Table 3 we report   the
fraction of  AGN with  luminosity larger than  $L_H=10^{42} erg s^{-1}$
 and the fraction of AGN with luminosity higher than
$L_H=10^{43} erg s^{-1}$, hosted by galaxies with  rest-frame magnitude brighter
than $M_R =-20$. When no AGN are identified  the upper limits are 
evaluated using the low number statistics
estimators by  Gehrels (1986). 
Overall, we find  an average  fraction  of $6.3\%$ for  AGN  with  $L_H> 10^{42} erg s^{-1}$
with a very large range (from less than $5\%$ to $22\%$). For the most luminous
AGN with  $L_H> 10^{43} erg s^{-1}$, we find  a global fraction of $2.1 \%$.
In Figure 4 we plot these individual fractions (for  $L_H > 10^{42} erg s^{-1}$
) or upper limits for our groups and small clusters (as green
symbols).
\begin{figure*}
\includegraphics[width=9.5cm,clip=]{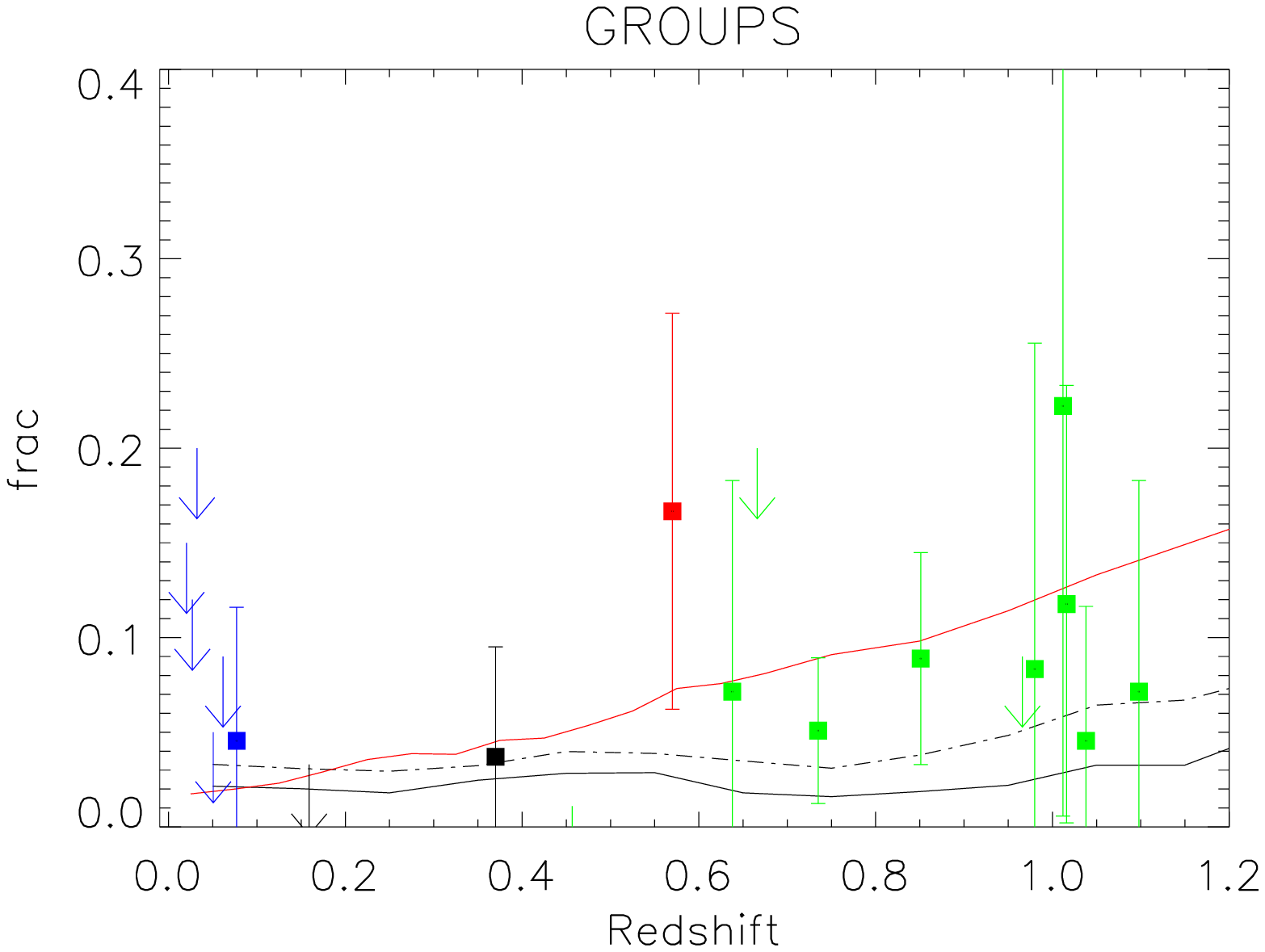}
\includegraphics[width=9.5cm,clip=]{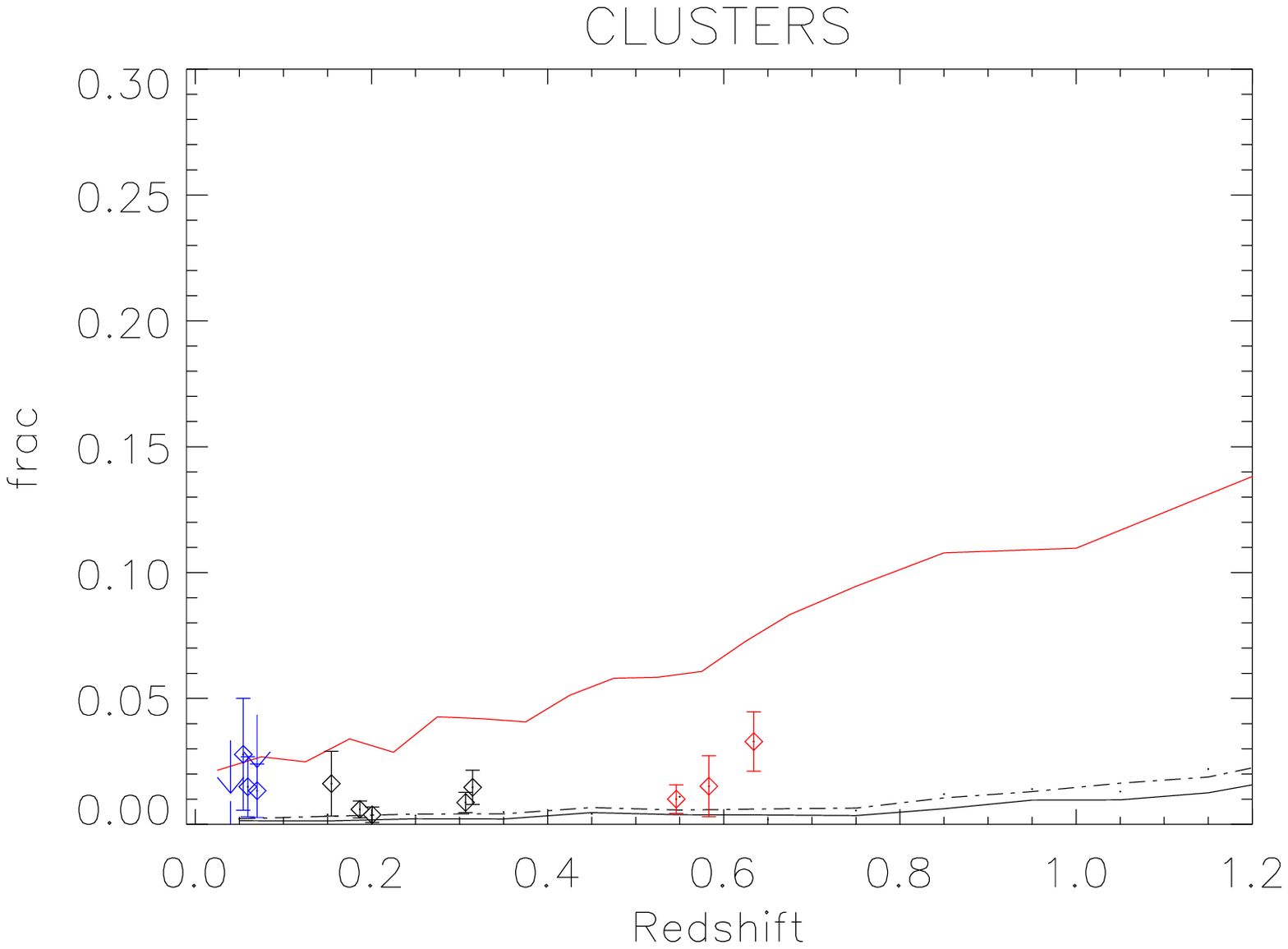}
\caption{Left  panel: the fraction of AGN with $L_H> 10^{42} erg s^{-1}$ in
  groups and small clusters with velocity dispersion $350 < \sigma < 700\ km
  s^{-1}$. Green symbols are structures from the present work, blue symbols
  are from Arnold et al. (2009), black symbols from Martini et al. (2009), red
  symbol from Eastman et al. (2007). Right panel: the same for clusters with
  $\sigma > 700\ km s^{-1}$. In both plots, to determine the  upper limits and to estimate the
    uncertainties on the fractions  we used  the low number statistics
estimators (at 1 $\sigma$) by  Gehrels (1986). 
 The red solid lines are the predictions from the
  Millenium simulation (Guo et al. 2011), the black solid and dash-dotted lines are
the nominal and maximal predictions using the  \citet{Menci2006} model, see
text for more details.}
\label{fig:evolution}
\end{figure*}
\subsection{The dependence of AGN fractions on redshift and velocity dispersion}
Our  results  can be immediately compared to the analogous  
analysis  of low redshift groups and  clusters by \cite{Arnold2009}.
They selected structures with  a range of velocity
dispersions (and richness) similar to ours and   extended   to groups with
$\sigma$s as low as 250 km/s and, on the other side, to few 
more massive clusters with $\sigma$ up to  900$ km s^{-1}$.
For a more accurate comparison we restrict their study to the same range of 
velocity dispersion probed by our sample, which is approximately 
between $350$ and
700$km/s$, thus including  six of their structures.
The result is a fraction of AGN  with $L_H > 10^{42} erg s^{-1}$  of  $\sim 1 \%$ at an average
redshift z=0.045. No AGN brighter than $L_H=10^{43} erg s^{-1}$ are hosted by groups
and small clusters in the local universe  in the sample of Arnold et
al. (2009), implying a limit of $<0.9\%$. 
In Figure  4 we report the values derived by Arnold et al. for these six 
groups  (represented as blue symbols); we  also include few more relevant
results from the literature, in particular the small clusters presented 
in Martini et al. (2009) 
at slightly higher redshift (Abell 1240 at $z=0.159$ and MS1512 at $z=0.37$,
black symbols) and  one of the structures studied in Eastman et al (2007) 
(Abell 0848 at z=0.67, red symbol), 
with a velocity dispersion that is within  our range. 
The  trend for increasing AGN fraction with redshift  is  clear: most of
the low redshift groups  have no AGN (and are plotted  as upper limits), while
at $z > 0.5$ many  have $f_A \sim 5-10\%$  amongst bright galaxies.
Note that in some cases, the luminosity of AGN in the above papers 
was  reported in different rest-frame bands: we convert it  to
2-10 KeV rest-frame, always assuming that the spectrum is represented by a 
power-law with photon  index $\Gamma=1.8$ as above. 
\\
The same trend we observe in groups/small clusters 
has  already been noted  in more massive clusters:  Eastman et
al. (2007) compared the 
AGN content in  clusters at z$\sim$0.6-0,7 to the analogous structures in
the local Universe  analysed by Martini et al. (2007) and 
found a factor of 10 increase. 
In Figure 4 (right panel) 
we also plot a collection of results from the literature on more 
massive structures (i.e. clusters  with  $\sigma > 700 km s^{-1}$): 
these include the three more massive clusters in Eastman et al. (2007) at
z$\sim$0.6-0.7 (red symbols), the
low redshift structures with  $\sigma > 700 km/s$ from Arnold et al. (2009)
(blue symbols) and the intermediate  redshift clusters analysed by  Martini et al. (2006) (black symbols). 
Although several  results have been
published on  massive clusters at redshift above 0.7, we do  not include them
in this plot   mainly because the available X-ray observations are not
sensitive to AGN with luminosities of  $L_X=10^{42}$  at these very high redshifts.
We remind  that for AGN with $L> L_H = 10^{43} erg s^{-1}$ in clusters, 
 Martini et al. (2009)  found  a considerable
evolution  from 0.2\% at $z<0.3$ to 1.2 \% at $z\sim 1$. 
\\
We conclude that groups behave like their more massive counterparts, 
in terms of AGN content and its evolution with time, and there 
is a net trend for an increasing AGN fraction hosted by galaxies 
brighter than a fixed limit ($M_R=-20$ in our case). 

From a comparison between the two panels of Figure 4 we see that 
groups contain  comparatively many more AGN that more massive clusters.
To test if the fraction of AGN  
depends significantly on the velocity dispersion of the
systems at a fixed redshift, we  run  a Spearman rank correlation: we first
apply the test to our own sample  
and   the result is a rank  coefficient r=-0.58 with
a probability of no correlation of P=0.06. So there  are indications 
of some anti-correlation between the velocity dispersion of a structure and
its AGN fraction, although with a large scatter. 
We then add  the  four structures studied by Eastman et
al. (2007) at z$\sim$ 0.6 which include 
three higher velocity dispersion systems (see above).
We repeated the Spearman  rank correlation test  with the total sample of 
15 groups and clusters and found a higher coefficient (r=-0.64) with a much higher 
significance (P=0.010). We therefore conclude that, at a given redshift, the 
 lower dispersion systems have comparatively more AGN at a fixed luminosity
 threshold, compared to the more massive structures.

\subsection{The AGN spatial and velocity distribution within groups}
The distribution of the AGN within  the clusters and groups in terms of
spatial position and relative velocity,   
can potentially offer clues on  the triggering of the active phase, its
     lifetime, and the fueling mechanisms.
If AGN are mainly fueled by galaxy-galaxy interactions, 
one expects that they  should be more prevalent in the outskirts 
of clusters/groups.  If gas-rich mergers are the primary mechanism for activating and fueling AGN, one expects higher
AGN fractions in environments where galaxies have an abundant supply of gas:
in this case  galaxies in the centers of rich clusters
should host less AGN since there is proportionally less cold gas (e.g. Giovanelli \& Haynes
1985).
However, a significant fraction of early type galaxies, 
which tend to lie in the centers of richest clusters, 
are known to harbour   AGN and LINERs. A relation between AGN and early-type galaxies could dilute 
or even reverse the trends predicted by gas-rich mergers or galaxy 
harassment. 
A further effect that can trigger AGN is the interaction 
with the central brightest cluster galaxy, which is itself often a
powerful AGN (e.g. Ruderman \& Ebeling 2005).  
The  relative importance of all these  effects could also vary from very
massive structures (where the velocity  differences are more marked) 
to groups and smaller clusters.

Martini et al. (2002) were amongst the first to  study the spatial 
distribution of X-ray selected AGN in clusters of galaxies at z
$\sim$0.06-0.31 and  found that the AGN with $L_X >
10^{42} erg s^{-1}$ and  $M_R <
-20$  were located more centrally compared to  inactive galaxies, although
they had comparable velocity and substructure
distributions to other cluster members.
Ruderman \& Ebeling (2005) studied the spatial distribution of 
X-ray point sources in 51 massive galaxy clusters at 
$0.3 < z < 0.7$, and concluded that they lie 
predominantly in the central 0.5 Mpc.
Similarly Martel et al. (2007) showed that the surface density
of the X-ray sources in  five massive  X-ray clusters at z$\sim 0.8-1.2$ 
is highest in the inner  regions  and  relatively
flat at larger radii, although AGN tend to avoid the very inner 
cores of clusters, i.e. regions of $\sim 200 kpc$.
The same was found by Galametz et al. (2009) for bright X-ray AGN
 for $0.5 < z < 1.5$ galaxy clusters and by  Bignamini et al. (2008) for
 RCS clusters at  z$\sim 0.6-1$ also showing a significant excess
 of medium luminosity X-ray AGN close to the centroid of the X–ray emission. 

\begin{figure}
\includegraphics[width=9cm,clip=]{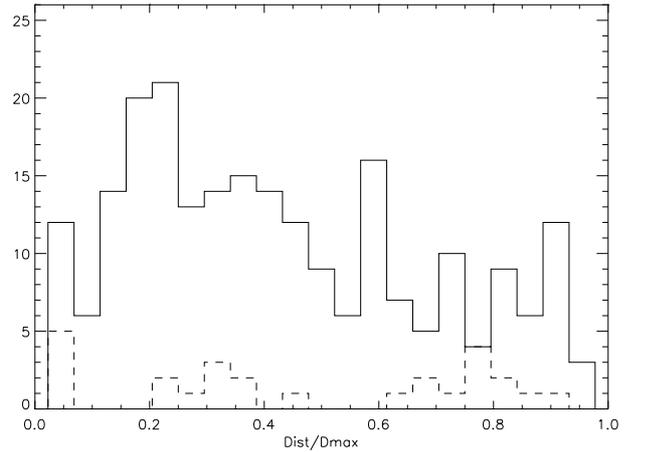}
\caption{The solid line shows the radial distribution of 
all galaxies brighter than $M_R=-20$ in
  the 11 structures (normalized by the maximum radius of each structure); the 
 dashed line is the same distribution for  AGN}
\label{fig:fig1}
\end{figure}
At variance with the above works, Gilmour et al. (2009) 
analysed  a sample of 148 galaxy clusters at 
$0.1 < z < 0.9$ finding 
that the  X-ray sources  are  quite evenly
distributed over the central 1 Mpc, while  Johnson et al. (2003) found that in the z$=$0.83 cluster
$MS 1054−0321$, the excess of X-ray  AGN is  at much 
larger radial distances, suggesting that they may  be associated
 with infalling galaxies. Finally we mention the recent work of Fassbender et al. (2012)
 in high redshift massive clusters, 
 indicated significant excess of low luminosity AGN in the inner (1Mpc)
 regions
as well as an excess of brighter soft band sources at much larger distances 
suggesting perhaps the idea of two different AGN populations and triggering  mechanisms of
nuclear activity.

A big caveat to the above studies  (with the exception of Martini et al. 
2002, 2007) is the lack the spectroscopic redshift confirmation for most  
or all X-ray AGN.
Moreover  a lot of the analysis reported are limited to very luminous AGN:
testing the distribution of more ``normal'' AGN can probe whether 
the AGN activity is more related to the host galaxy properties, or to the 
environment. We  therefore analysed the spatial distribution of active and inactive 
galaxies in our structures;   
we used  as cluster/group centers the position given by the search algorithm, 
unless a clearer center is given by 
the presence of extended X-ray emission (as in the case of GS 5) 
or by the position of a dominant  brightest  galaxy.
We then determined the distance of the AGN and inactive galaxies 
from  the center and normalized it by the extent of each system. 
The resulting distribution for normal and active galaxies is presented in
Figure 5.
We see no indication
for a concentration of AGN towards the cluster/group center compared to 
the entire galaxy population. The distribution of AGN is 
actually flatter than that of the underling population, i.e. 
there are comparatively more AGN in
the outer parts of the structures.
To determine whether the AGN sample is  consistent with being randomly
drawn from the parent sample of galaxies or not, we run a non parametric K-S test. 
We find that the probability of this event is very 
low P=0.055, so most probably the AGN
are distributed differently from the underling  global population.
Our conclusion is therefore that moderately luminous AGN tend to
preferentially  reside in the outskirts of structures compared to normal galaxies.
\\
One possibility is that  these AGN might have just entered in the
cluster/group potential: in this case we also expect that they would be on more
radial orbits compared to the rest of the population.
Following, e.g., Martini et al. (2009) we determine the cumulative velocity
distribution for all  AGN, normalised by the cluster velocity dispersion in
each case ($v-v_c/\sigma$). We find that the distribution agrees well with a
Gaussian, thus there is no evidence that the AGN have a larger velocity
dispersion than the rest of inactive galaxies.

In conclusion we find that our AGN are preferentially located in
 the outskirts of the  structures
but have the same velocity distribution as the rest of the galaxy population.
This would support to the idea  that mergers and  tidal interactions
are one of the main instigators of AGN activity; AGN are preferentially located
in  intermediate
density regions  (outskirts of groups and clusters) which 
are the most conducive to galaxy-galaxy interactions
because of the elevated densities, compared to the field,
but the relatively low velocities 
compared to cluster cores.
However given the many discrepant results in the literature, 
this scenario  has to be tested further with larger, high redshift group samples. 
\subsection{The color-magnitude relation of AGN in dense environment}
It has been proposed that AGN may be responsible for the moderation of
star-formation activity, either by sweeping up the gas from the galaxy thus
stripping star-formation, or by inhibiting further gas from cooling and
infalling (e.g., Maiolino et al. 2012, Croton et al. 2006).
In this context one can predict the AGN hosts to be located in distinct 
regions of  the color-magnitude diagram for galaxies.
In particular the  color distribution  of AGN compared to those of 
the general (inactive) galaxy population can place constraints on the relative
timing of the physical processes that take place in the galaxies:
for example, if the nuclear  activity timescale  is longer than the 
timescales on which star formation activity is quenched, or if there are
dynamical delays between star-burst and AGN activity in galaxy nuclei,
AGN hosts will tend to be preferentially red compared to the general inactive
galaxy population.

We therefore investigated the colors of our AGN host galaxies
compared to the underling galaxy population: 
we remark that our AGN are all of modest luminosities 
hence we expect that their  optical light is   dominated by host galaxy
contribution  and not influenced in a significant way by the AGN, 
therefore the colors we determine correspond to the stellar
population.
 We further checked this issue by  exploiting  the fitting made 
by Santini et al. (2012) for X-ray sources in Goods-North and South. Here the
spectral energy distribution (SED) of galaxies hosting Xray sources 
was fitted with a double component, one for the AGN and  one for the
stars (see for example Figure 2 of that paper, for two cases, 
a type 1 and a type 2 AGN). 
As a result  we get for the best-fit solution the relative
contribution to the total luminosity of the two components at a rest-frame
wavelength 6500 \AA (R-band). We have verified that  for our sources 
the contribution of the AGN 
component is not significant  in all cases.

 The color-magnitude diagram for the X-ray sources and of the general 
 galaxy population is  shown in Figure 6: the galaxies  show the
 well-established bimodality of colors at this redshift, while it is 
clear that X-ray sources are not randomly distributed over the same
region as the galaxies. 
All AGN hosts have colors redder than $U-B>0.5$ and tend to  
reside mostly in the green valley, on the red sequence or the top
of the blue cloud. 

This plot can be immediately compared to an analogous one by Nandra et
al. (2007, Figure 1 of their  paper) 
who analysed the Color-Magnitude Relation for X-Ray selected  AGN
in the AEGIS field at a similar redshift 
($0.6 < z < 1.4$). If we neglect the brightest of their AGN, which are
actually QSOs and have very blue colors, 
 we see that in their  case  AGN tend to populate the entire    
 color magnitude diagram;  there are also AGN in the blue
cloud, although they are a relative minority. The fraction of galaxies which
are also X-ray sources in the red sequence, green valley and blue cloud are
3.4, 4.2 and 0.9\% respectively.
 Silverman et al. (2008) also showed that 
the fraction of galaxies hosting AGN peaks in the "green valley" ($0.5<
U-V<1.0$) especially in the presence of large scale structures. 
They further showed that at $z > 0.8$, a distinct, blue population of host
AGN galaxies is prevalent, with colors similar to the star-forming galaxies. 
More recently, Rumbaugh et al. (2012) confirmed that in clusters and
superclusters  many 
AGN are located in the green valley, consistent with being a transition 
population.

From the comparison of the  color-magnitude diagram  of AGN 
in groups/clusters (our work Figure 6) with the CMD of AGN in the field 
(Nandra et al., Silverman et al.)
 we can see that in groups/cluster the AGN basically avoid the blue cloud, while in the field, AGN are also
present in the blue cloud.
If merger-induced AGN activity is   associated with the process
that quenches star  formation in massive galaxies (e.g. di Matteo et
al. 2005), causing the migration of
blue cloud galaxies to the red sequence (Croton et al. 2006; Hopkins et
al. 2006b), then the different color-distribution  of AGN in the field 
and in groups indicates that  these phenomena are   more rapid in dense
environments. Galaxies hosting  AGN  abandon the blue cloud 
more rapidly in clusters and groups,  as inferred from  our data, compared to
what happens in the field.

\begin{figure}
\includegraphics[width=11cm,clip=]{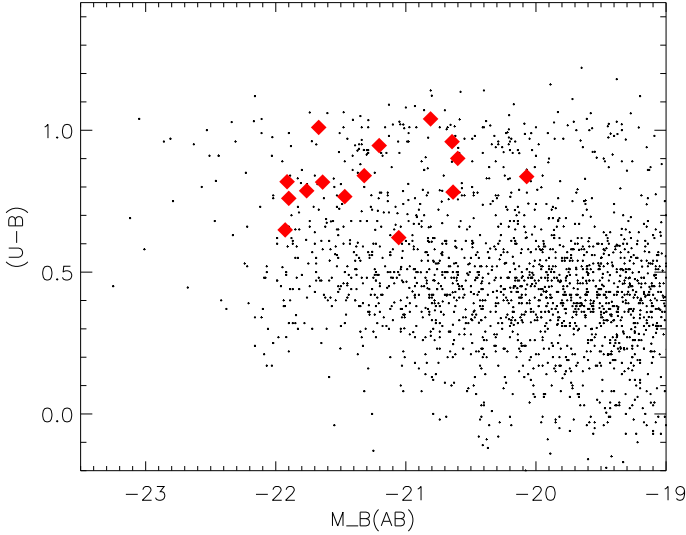}
\caption{The color magnitude diagram for all galaxies in  the clusters and
  groups (small black dots) with the positions of the AGN marked by the red diamonds.}
\label{fig:cdm}
\end{figure}

\begin{figure}
\includegraphics[width=10cm,clip=]{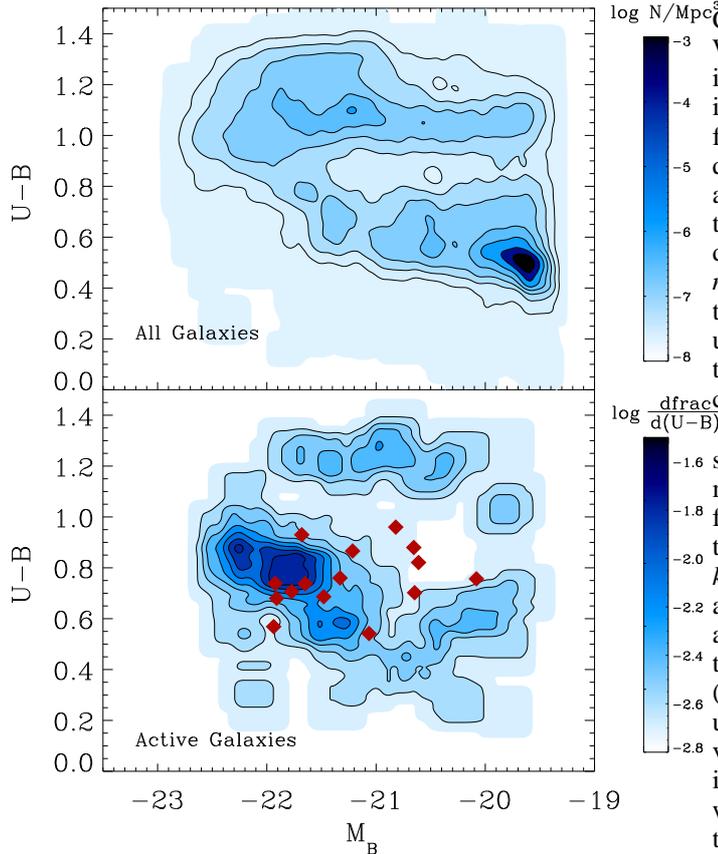}
\caption{ The predicted  color magnitude relation for all galaxies (top panel) and
  active galaxies (lower panel) with a hard X-ray luminosity larger than
  $10^{42} erg s^{-1}$ from the M04 model: contours are number densities.
In the lower panel we overplot the AGN
  observed in our sample as red diamonds. }
\label{fig:cdmmodel}
\end{figure}

\section{Comparison to  model predictions}

A comparison between the observed results and the 
predictions of semi-analytic models (SAM) that include AGN growth, can help us
understand what are the main physical processes that drive the formation and the
fueling of black holes.
In the previous section we have derived that the frequency and colors 
of AGN depend quite strongly on the environmental density, with marked differences
between field, groups and massive clusters. We will therefore compare our
results to models that 
analyse the processes of  AGN triggering and fuelling within a fully cosmological framework.
Broadly,  there are two main modes of AGN growth in these models:  
the so called ``radio mode'' and the ``quasars mode''.
The quasar mode applies to black hole growth during gas-rich mergers  where
 the central black hole of the major progenitor grows both by 
absorbing  the central black hole of the minor progenitor and by accreting the
cold gas. 
In the radio mode, quiescent hot gas is accreted onto the central super-massive
black hole; this accretion comes from the surrounding hot halo and is
typically well below the Eddington rate. This model captures the mean behaviour of the black
hole over timescales much longer than the duty cycle.

We will employ two different semi-analytic models, one that
implements  only the quasar mode and one that implements both.   
The  model of Menci et al. (2004 M04 in the following) falls in the first
category and is particularly tailored to follow the evolution of AGN. In this model 
the accretion of gas in the central black holes, 
is triggered by galaxy encounters, not necessarily leading to bound mergers, in common
host structures such as clusters and especially groups;
these events destabilize part of the galactic cold gas and hence
feed the central BH, following the physical modeling developed
by Cavaliere \& Vittorini (2000). The
amount of cold gas available, the interaction rates, and the
properties of the host galaxies are derived as in  Menci et al. (2002).
As a result, at high redshift the proto-galaxies grow rapidly by
hierarchical merging; meanwhile   fresh gas is imported
and the BHs are fueled at their full
Eddington rates. At lower redshift, the dominant dynamical events
are galaxy encounters in hierarchically growing groups; at this point 
refueling diminishes as the residual gas is exhausted, and the
destabilizing encounters also decrease.
This model successfully reproduces the observed properties of both galaxies
and AGN across a wide redshift range (e.g. Fontana et al. 2006; Menci et
al. 2008b; Calura \& Menci 2009; Lamastra et al. 2010).
\\
We further  compare our results to the output of a SAM model implemented in
the Milleniumn simulations (MS in the following) as in Guo et al. (2011). 
For black hole growth  and AGN feedback they 
follow Croton et al. (2006), who implement both  ‘quasar’ mode and ‘radio’
mode. In the ``quasars  mode'' black hole accretion is allowed 
during both major  and minor mergers, but  the efficiency in the latter is
lower because the  mass accreted during a merger depends, among the
  other factors, on the ratio   $m_{sat}/m_{central}$ (eq. 8 in Croton et al. 2006). 
In the ``radio mode'', the growth of the super-massive black hole 
is the result of continuous hot gas accretion
 once a static hot halo has formed around the host galaxy of the black hole. 
This accretion is assumed to be continual and quiescent
 (see Croton et al. 2006 for more
details).

From the two SAMs, we select  all galaxies residing in massive halos
(on the scale of groups and clusters), with rest-frame magnitudes brighter
than $M_R =-20$ as in our observations.
As for the real clusters and groups,  we divide the simulated 
structures into those  with a velocity
dispersion between 400 and 700 $km s^{-1}$ (i.e., groups and small clusters)
and those  with sigma above 700 $km s^{-1}$ (massive clusters). 
From the simulations we actually know the total mass of the
corresponding dark matter halos, which is related to the velocity 
dispersion via  ${v_c} ^3 =(M / f(z) )*(h/0.235)$, where  $f(z) = H(z)/H_0$, 
with halo mass in unity of $10^{12} M\odot$
and v is in unity of 100  $km s^{-1}$.  
The SAMs provide the total bolometric luminosity of each AGN; 
to convert this into  observed X-ray luminosity in the 2-10keV rest-frame band, 
we follow  the relations found by  Marconi et.
al. (2004) applied to our luminosity limits ($L_H > 10^{42}$ and $L_H> 10^{43}$)
\\
For all galaxies, the model computes  the total stellar mass (M*): 
at each redshift we determine  the mass corresponding to $M_R=-20$ from the 
relation between stellar mass and $M_R$ derived
from the GOODS-South catalog (Grazian et al. 2006). We then use this mass 
to select mock galaxies brighter than  $M_R=-20$.
Since the mass-magnitude   relation  has a scatter we make two different
 predictions. In one case  we use the best fit value of the 
mass-magnitude relation to determine $M_R$ and then 
 select galaxies  
(filled curve in Figure 4, nominal prediction). 
In the second case we use the maximum  stellar mass M* corresponding to
$M_R=-20$  as a selection threshold. 
 In this second way  we select
 more massive galaxies and therefore the  probability to find an  AGN in the
 galaxies is higher. This is the  upper envelope of our prediction (dashed
 curve in Figure 4, maximal prediction).The MS  model gives directly the
   R magnitude  of the mock galaxies so for this
  model we have only the nominal prediction.
The resulting fractions of AGN with $L_H > 10^{42}$  in groups and
clusters hosted by galaxies  brighter
than $M_R =-20$ found in  the two models are presented in Figure 4, along with  the
observed data.
\\
 The MS  model
tends to over-predict the fraction of AGN, especially for
massive structures and at high redshift, while is it more in agreement with the data for 
groups. It also predicts a very marked increase of the  AGN fraction  with
redshift, more
pronounced than what is observed in the data. This steep increment  is
  linked to the marked rise of major mergers (the only
mergers considered for the quasar mode) towards high redshift.
This  model 
predicts a modest dependence of the AGN fraction on the velocity
dispersion of the systems: for example at z$\sim 0.6$ simulated  groups
contain  only $\sim$20\% more AGN than the more massive structures,
while the observed difference is much larger.
\\
The M04 model predicts a milder increase of AGN fraction with redshift,
both for massive and smaller systems: this is due to the fact that in
this model  minor mergers and close encounters are also very 
important  and their
frequency does not depend so strongly on redshift, since the small Dark Matter
halos  continue to merge frequently until low redshift.
The M04 model tend  to under-predict slightly 
the observed AGN fractions at all redshifts: 
the observed offset between the data and the predictions is 
 approximately a factor of 3, both for clusters and for groups. This 
can be  explained by the known problems of semi-analytic model  
 that tend to overestimate the  number of galaxies at the faint end of the
 luminosity function.  
In particular for the M04  model this discrepancy at the faint end
 was extensively discussed in Salimbeni et al. (2008) and is clearly
observed at the magnitude limit that we are using in this study ($M_R=-20$).
\\
The  M04 model predicts a marked difference between groups and
clusters: for example at z$\sim$0.6 
 groups/small systems contain a factor of 5 more AGN compared
to massive clusters, in agreement with what  is observed on the data.
Indeed, in this model the fraction of 
gas accreted during mergers and
fly-by is  inversely proportional to the velocity dispersion of the structures, 
therefore  for clusters it is lower than in groups.
This effect is in addition to the increased merger rate between galaxies 
in groups, as compared to clusters, due to the lower encounter velocities
in these small systems. In this sense, the agreement between the observational and predicted
trends with velocity dispersion  and with redshift validates the implemented mode
of AGN growth in the M04 models.

We further check  if the models can reproduce  the colors of the AGN
in dense environments. To this aim, we find that the  MS model
has problem in reproducing the colors of the general galaxy population  
in clusters and groups. Guo et al. (2011) already remarked 
clear differences between SDSS observations and model predictions 
 in the slope of the red sequence and in the number of fainter red-sequence
 galaxies. The same was also noticed by de la Torre et al. (2011), who found
 that the De Lucia \& Blazoit (2007) implementation on the Millenium Simulation 
 does not reproduce quantitatively the observed intrinsic colour
 distributions of galaxies, with much fewer very blue
 galaxies  and many more “green valley” galaxies in the model than in the
 observations, at redshifts $0.2<z<2.1$. In addition, the model predicts an excess of red galaxies at low redshift.
We therefore decided to employ only the M04 model for this comparison: 
  this model  does a good job in reproducing 
the color bimodality of galaxies up to high redshift, as shown
 in the upper panel of Figure 7 where  
 we plot the predicted color magnitude
relation for all mock galaxies. The galaxies are located in a clear
red-sequence and blue cloud 
and are well  matched to the colors of the observed galaxies (Figure 6).  

In the lower panel we plot the predicted colors of
active galaxies which are selected as objects  with a total rest-frame
magnitude brighter than $M_R=-20$,  hosting an AGN with luminosity exceeding
$10^{42} erg s^{-1}$, and included in halos of mass comparable to our small clusters and groups.
Here we also plot the colors of our observed AGN.

 The U-B color range of the predicted AGN is well matched to the observations, 
most AGN having $0.5< U-B <1$, like the observed ones.
The model predicts the presence of a small fraction 
of extremely red AGN, that reside on top of the red sequence, i.e., 
that are even redder than the typical red-sequence galaxies. 
We do not observe these extremely red AGN but this might be just due to  lack 
of statistics. The model also predicts AGN in 
galaxies brighter than $M_UV=-22$ that we do not observe.
Again this could be due to lack of statistics, since these extremely luminous
galaxies are  quite rare in our observed sample (see Figure 6). Alternatively 
it might be  that mock galaxies hosting  AGN of $L\sim 10^{42} erg s^{-1}$ 
become too bright. Indeed in the M04 model 
each encounter/merger  that triggers AGN activity also
triggers  star-formation, thus enhancing  the UV luminosity of the host
galaxy; the relative proportion of gas that feeds AGN and star formation,
which is now fixed to 
approximately 1 to 4 (see Menci et al. 2006) might need to be revised.

\section{Summary and conclusions}
  We have explored the AGN content in small clusters and groups in the two
  GOODS fields, exploiting the ultra-deep 2 and 4 Msec Chandra
  data and the deep  multiwavelength observations  available.
  We have  used our previously tested cluster-finding algorithm to
identify structures, exploiting the available
spectroscopic redshifts as well as accurate photometric redshifts. We
identified 9 structures in GOODS-south (already presented in Salimbeni et
al. 2009) and 8 new structures in the GOODS-north field. 
To have a reliable estimate of AGN fraction, we restrict our study to  structures where at least
2/3 of the galaxies brighter than $M_R=-20 $ have a spectroscopic
redshift.
We identified those clusters members that 
coincide with X-ray sources in the  4 and 2 Msec source catalogs
(Luo et al. 2011 and Alexander et al. 2003 respectively), and
with a simple classification based on total rest-frame hard luminosity and 
hardness ratio we determined 
if the X-ray emission originates from AGN activity or it is related to
the galaxies'star-formation activity. We then computed the frequency of AGN
in each group: we found that at $z\sim 0.6-1.0$ the average fraction of 
AGN with $ Log L_H > 42$ in galaxies with $M_R < -20$ is $6.3 \pm 1.3 \%$, i.e. much
higher than the value found in lower redshift groups, which is just
1\%. This fraction is also more than double the fraction found in more massive
clusters at a similar redshift. 
We have then  explored the AGN spatial distribution within the structures and found that they tend to populate the outer
regions rather than the central cluster galaxies.
The colors of AGN in structures 
are  confined to the green valley and red-sequence, 
avoiding the blue-cloud, whereas in the field  AGN are also present  in the
blue cloud (e.g. Nandra et al. 2007). 
If the   AGN activity is   associated with the process
that quenches star  formation in massive galaxies (e.g. di Matteo et
al. 2005), causing the migration of
blue cloud galaxies to the red sequence (Croton et al. 2006; Hopkins et
al. 2006), we conclude   that  these phenomena are  more rapid in dense
environment compared to what happens in the field. 

We finally compared our results to the predictions of two sets of semi analytic
models: the M04 model (Menci et al. 2006) and one implemented on the Millenium
Simulation by Guo et al. (2011).
The MS model predicts a dependence of AGN content with redshift (both for
clusters and groups) that is much
steeper than what observed and a very modest
difference between massive and less massive structures.
The MS04 does a good
job in predicting the redshift dependence of the AGN fractions, and the 
 marked difference that is observed  between groups and massive clusters.
This agreement validates the implemented mode
of AGN growth in the model and in particular stresses
the importance of galaxy encounters, not
necessarily leading to mergers, as an efficient AGN triggering mechanism. 

The M04 model also reproduces accurately 
the range of observed AGN colors and  their position in the color-magnitude
diagram, although it tends to find AGN in galaxies that are on average
slightly more luminous than the observed ones. It also predicts the presence 
of a small fraction  of extremely red AGN, residing on top of 
the red sequence.
We do not observe these extremely red AGN but this might be due to  lack 
of statistics: we therefore plan to expand our analysis to other fields, 
with similar multiwavelength data and  deep X-ray observations 
to study the AGN content.
In particular we are currently working on the UDS field, thus more than 
doubling  the area (and the statistics) presented of this paper.
In this way we will be able to test, amongst other things,  
if the predicted extremely red AGN exist,  and we
will be able to place more stringent  constrains on the relative timing 
of AGN activity and the quenching of star formation at high redshift.

%

\end{document}